\documentclass[a4paper,12pt]{article}
\usepackage[format=hang]{caption}
\usepackage{epsfig,amssymb,amsmath,graphicx,subcaption,verbatim,hyperref,xcolor,
ulem,epstopdf,psfrag,pstool,braket,array,enumerate,multirow,wrapfig}
\usepackage{graphics,color}

\begin{document}
%

\newcommand{\be}{\begin{equation}}
\newcommand{\ee}{\end{equation}}
\newcommand{\bea}{\begin{eqnarray}}
\newcommand{\eea}{\end{eqnarray}}
\newcommand{\bean}{\begin{eqnarray*}}
\newcommand{\eean}{\end{eqnarray*}}
\font\upright=cmu10 scaled\magstep1
\font\sans=cmss12
\newcommand{\ssf}{\sans}
\newcommand{\stroke}{\vrule height8pt width0.4pt depth-0.1pt}
\newcommand{\Z}{\hbox{\upright\rlap{\ssf Z}\kern 2.7pt {\ssf Z}}}
\newcommand{\ZZ}{\Z\hskip -10pt \Z_2}
\newcommand{\C}{{\cal C}}
\newcommand{\R}{\hbox{\upright\rlap{I}\kern 1.7pt R}}
\newcommand{\HH}{\hbox{\upright\rlap{I}\kern 1.7pt H}}
\newcommand{\CP}{\hbox{\C{\upright\rlap{I}\kern 1.5pt P}}}
\newcommand{\identity}{{\upright\rlap{1}\kern 2.0pt 1}}
\newcommand{\half}{\frac{1}{2}}
\newcommand{\quart}{\frac{1}{4}}
\newcommand{\pr}{\partial}
\newcommand{\bm}{\boldmath}
\newcommand{\I}{{\cal I}} 
\newcommand{\M}{{\cal M}}
\newcommand{\N}{{\cal N}}
\newcommand{\e}{\varepsilon}
\newcommand{\nn}{\nonumber}

\thispagestyle{empty}
\rightline{DAMTP-2019-10}
\vskip 3em
\begin{center}
{{\bf \large The Inevitability of Sphalerons in Field Theory
}} 
\\[15mm]

{\bf \large N.~S. Manton\footnote{email: N.S.Manton@damtp.cam.ac.uk}} \\[20pt]

\vskip 1em
{\it 
Department of Applied Mathematics and Theoretical Physics,\\
University of Cambridge, \\
Wilberforce Road, Cambridge CB3 0WA, U.K.}
\vskip 12mm

\abstract{}

The topological structure of field theory often makes inevitable the existence
of stable and unstable localised solutions of the field
equations. These are minima and saddle points of the energy. Saddle point
solutions occurring this way are known as sphalerons, and the most
interesting one is in the electroweak theory of coupled 
$W$, $Z$ and Higgs bosons. The topological ideas underpinning
sphalerons are reviewed here.


\end{center}

\vskip 5em

\centerline{Based on Lecture at Royal Society Scientific Discussion Meeting:}
\centerline{Topological Avatars of New Physics, 4-5 March 2019} 

\vskip 9em

\leftline{Keywords: Sphaleron, Topology, Saddle Point, Electroweak Theory}
\vskip 1em

\vfill
\newpage
\setcounter{page}{1}
\renewcommand{\thefootnote}{\arabic{footnote}}


\section{Idea of a Sphaleron} 
\vskip 3mm

A sphaleron is a static, finite-energy solution of classical field 
equations that is unstable \cite{KlMa}. The origin of the word is from ancient
Greek: $\sigma\phi\alpha\lambda\epsilon\rho${\small{\textit o}}$\varsigma 
=$ unstable, ready to fall. A sphaleron is a stationary 
point of the energy, but not a minimum. It is analogous to a transition state
in molecular physics -- a saddle point in the configuration space of 
atomic locations. The field theories we consider here are usually Lorentz
invariant. Although the sphaleron solution only depends directly on the energy
functional for static fields, the way the sphaleron can be produced and
decay will depend on the full, time-dependent field equations.

Topological structure in a field theory, together with the
scaling properties of the energy, can make a sphaleron's existence inevitable,
although a rigorous proof of this may not be available.
Like a topological soliton, a sphaleron is localised in space, and has no
translation symmetries, but it sometimes has the maximally allowed 
rotational symmetry.

Sphalerons can occur in theories without finite-energy solitons, 
e.g. in the electroweak theory. Sphalerons also occur in theories 
with solitons, e.g. as monopole-antimonopole pairs \cite{Tau}, and 
in the Skyrme model.

In a given field theory, let $\C$ denote the configuration space of
static fields satisfying vacuum boundary conditions and
having finite energy. $\C$ is an infinite-dimensional manifold.
In gauge theories, we define $\C$ as the space of fields with gauge 
transformations quotiented out as far as possible. This can be achieved by
judicious gauge fixing.

$\C$ may be connected, or maybe not. In theories with topological 
solitons, $\C$ is disconnected. In the simplest and best known cases,
the connected components $\C_n$ are labelled by an integer soliton
number $n$, called the topological charge. Usually there is a
(topological) charge reflection symmetry, so antisolitons with $n$
negative have the same energy as solitons with $n$ positive.

In physical theories, the energy $E$ is bounded below and has a minimum on 
$\C$ -- the vacuum solution -- but it has no maximum. If $\C$ is disconnected,
the vacuum is in $\C_0$, and the energy minimum in $\C_1$ is the 
stable 1-soliton. Examples of solitons are monopoles and
Skyrmions in 3D (three spatial dimensions), vortices, baby Skyrmions and
sigma-model lumps in 2D, and kinks in 1D \cite{book,Shn}. There 
are many variants of these types of soliton.

In sectors with higher soliton numbers, $n \ge 2$, it is not so obvious
that $E$ attains its minimum. There is a danger that it is
energetically favourable for a putative $n$-soliton to break up into 
clusters with soliton numbers $n-n'$ and $n'$, for some $n'$. If 
soliton clusters attract, then it is expected that the $n$-soliton
exists as a stable, minimal energy solution.   

The various possibilities are illustrated by abelian Higgs vortices in
2D, where there is a dimensionless coupling parameter $\lambda$ 
\cite{JR}. For $\lambda < 1$ vortices attract and merge, and there is a stable, 
circularly symmetric $n$-vortex solution for all $n$; for $\lambda > 1$ 
vortices repel, and there is no energy minimum for $n > 1$. But the 
circularly symmetric solution of merged vortices persists as an
unstable, static solution of the field equations, i.e. as a
sphaleron. It is unstable to break-up into $n$ individual vortices that
drift off to infinity. Finally, at critical coupling, $\lambda = 1$, 
there is a whole moduli space of static $n$-vortex solutions, with dimension
$2n$. These are all energy minima, having the same energy. The moduli
are the $n$ independent vortex locations \cite{JT}.

\vskip 2mm

\section{Morse Theory}
\vskip 3mm

Morse theory is a basic tool relating the topology of a manifold 
$M$ to the stationary points of a function $f$ on $M$ \cite{Mi}. The theory is 
simplest for $M$ finite-dimensional, compact, and connected, and with
$f$ a function of generic type, meaning that its stationary points are
isolated and that the matrix of second derivatives at each stationary
point (the Hessian) has only positive and/or negative eigenvalues (and no
zero eigenvalues).  

For such a function on a two-dimensional, compact, connected surface $M$, 
there are three types of stationary point: (local) minima, saddle
points and (local) maxima, where the Hessian has, respectively, zero, one or
two negative eigenvalues. The numbers (denoted $\sharp$) of such 
stationary points must satisfy the equality
\bea
&& \sharp({\rm minima}) - \sharp({\rm saddle} \, {\rm points}) 
+ \sharp({\rm maxima}) \nonumber \\
&& \qquad = \, {\rm Euler} \, {\rm number} \, {\rm of} \, M \, 
= \, 2(1 - {\rm genus}) \,,
\eea
and there are further (Morse) inequalities that we will not discuss.

For example, on a 2-sphere (genus 0), there may be just 1 minimum and 
1 maximum and no saddles, but saddles are inevitable if 
$\sharp({\rm minima}) > 1$ or $\sharp({\rm maxima}) > 1$.
As an illustrative function, consider the surface of a 
cube, which is topologically a sphere. The distance from the centre of
the cube is a function of direction, and hence a function on the
sphere. It has 6 minima at the face centres, 12 saddles at the edge
mid-points, and 8 maxima at the vertices. Note that $6 - 12 + 8 = 2$,
as expected. Saddle points are also inevitable for any function on a 
surface whose genus is greater than 0 (e.g. a torus with genus 1).

On a compact manifold $M$ of any dimension, one can find saddle points of $f$ 
by connecting a minimum to another minimum by trial
paths. Along any such path there will be a point where $f$ has its
maximum value, and then one can seek to minimize this maximum value over all 
paths. The minmax point exists because of the compactness of the
surface, and is a saddle of $f$ with one unstable direction. This
construction is due to Ljusternik and Schnirelman \cite{Lj}.

The maximum along a single, well-chosen path can give a good 
estimate for the position of the saddle and an upper bound for 
the value of $f$ at the saddle.

A variant of this approach works if $f$ has just one minimum, but 
there are non-contractible loops on $M$, starting and ending at the 
minimum. Now one finds the point on
each loop where $f$ has its maximum, and applies a minmax search to
all loops in a particular topological class (homotopy class) of
non-contractible loops. This works, e.g. on a torus, and determines a 
saddle point of $f$. 

\vspace{2mm}

\section{Saddle Points in Field Theory}
\vspace{3mm}

The method of non-contractible loops can be applied to field theories.
Let us assume the minimum of the energy $E$ is unique. The minmax field 
configuration on a class of loops through the minimum is a saddle
point of $E$, i.e. a sphaleron with one unstable
mode. Higher-dimensional, non-contractible spheres of fields can be 
used to find higher-energy saddles with more unstable modes.

This method is not foolproof, because field configuration space is
infinite-dimensional and lacks an obvious notion of compactness. Field
energy can drift off to infinity in various ways \cite{Tau}. In particular, the 
method fails when non-contractible loops exist but spatial rescaling 
of the fields can reduce the minmax energy to zero. For example, in 
pure 3D Yang--Mills gauge theory, and 1D sigma models, there are 
non-contractible loops related to instantons (topologically
non-trivial solutions of the field equations in 4D and 2D, 
respectively), but the maximum of the energy along such a loop can be
made arbitrarily small by a length rescaling.

It is also important, when applying the method in a gauge theory, to 
avoid the use of a non-contractible loop of gauge transformations,
along which the energy is simply constant. Morse theory na\"ively fails
in gauge theory because the Hessian at a stationary point of the energy 
has infinitely many zero eigenvalues associated with infinitesimal
gauge transformations. To avoid this degeneracy it is necessary to 
fix the gauge.

\vspace{2mm}

\section{Saddle Points in Gauge/Higgs Theory}
\vspace{3mm}

In gauge theories with Higgs symmetry breaking one can overcome these
difficulties. In a 3D gauge theory with Higgs field $\Phi$, the energy 
is of the generic form
\be
E = \int_{\R^3} \left( |F^2| + |D\Phi|^2 + V(|\Phi|)\right) \, d^3x \,, 
\ee
where $F$ is the field tensor constructed from the gauge potential
$A$, and $D\Phi$ is the gauge-covariant derivative of $\Phi$.
Terms scale in opposite ways under a length rescaling ${\bf x} \to
\mu {\bf x}$, so an energy minimum occurs at a finite length scale. 

An effective gauge fixing is to impose the radial gauge 
condition on the gauge potential, $A_r = 0$. This can be done continuously
for any continuous family of gauge and Higgs fields. It depends on choosing 
an origin, but that is not a problem for energetically localised
field configurations. 

Superficially, the radial gauge choice allows further gauge transformations 
depending only on the angular coordinates, but such gauge
transformations would generally be singular (multi-valued) at the 
origin. The only remaining gauge freedom is by rigid (global) gauge 
transformations. Such global gauge transformations can also
be partially suppressed by a base point condition -- for example,
requiring the Higgs field to take a specified value as one moves off to
infinity in a particular direction. Any residual global gauge freedom
is usually easy to deal with in the topological discussion.

In the following, we restrict our attention to two well-known 
examples of gauge/Higgs theory. The first is the Georgi--Glashow
model, with gauge group SU(2) and a real, adjoint (triplet) Higgs
field $\Phi$ \cite{GG1}. The standard quartic Higgs potential requires that 
the vacuum Higgs field lies on the 2-sphere $|\Phi| = 1$. The Higgs 
vacuum manifold is therefore $S^2$. The second is the bosonic
truncation of the electroweak theory of Weinberg and Salam. Here the
gauge group is U(2), with the SU(2) and U(1) coupling constants having
independent values (their ratio defines the weak mixing
angle). The Higgs field is a complex doublet $\phi$, and again the
quartic Higgs potential requires the vacuum Higgs field to satisfy
$|\phi| = 1$. The Higgs vacuum manifold is therefore $S^3$. 
(Here we have normalised the Higgs fields to absorb the vacuum 
expectation values. Note that the Higgs vacuum manifold is not a
point, because we have fixed the gauge using the gauge potential,
and not the Higgs field.) 

Finite-energy field configurations have a Higgs field that at spatial
infinity takes values on the Higgs vacuum manifold. By imposing
the radial gauge condition, $A_r = 0$, we ensure that the value of the
Higgs field on the sphere at infinity is well defined (because
the finite-energy condition requires the radial covariant derivative
of the Higgs field to decay rapidly). A field configuration is therefore
characterised by a mapping from the sphere at infinity to the Higgs
vacuum manifold, i.e. by a map $\Phi_{\infty}: S^2 \to S^2$ in the Georgi--Glashow
model, and by a map $\phi_{\infty}: S^2 \to S^3$ in the electroweak theory. These
maps are well defined up to a global gauge transformation that acts by
a global rotation on the target.

There is no further topological information in a field configuration,
because in both models the Higgs field is linear and not subject to any
constraint. The gauge potential in also essentially linear, so
unconstrained in finite regions of space. The gauge potential and Higgs
field are correlated at infinity (because angular covariant
derivatives of the Higgs field vanish at infinity), but all the
topological information about a field configuration, and more
importantly, about any continuous family of field configurations, is captured
by the family of maps from $S^2$ to the Higgs vacuum manifold ($S^2$
or $S^3$ in our examples). 

In the Georgi--Glashow model, the space of these maps is disconnected. 
A map $\Phi_{\infty}: S^2 \to S^2$ has a degree $n$, and the components
of the field configuration space are denoted $\C_n$. Physically, $n$ is the
monopole number. This is because, in $\C_n$, the asympototic gauge field has the
character of a magnetic monopole field, with magnetic charge
proportional to $n$. The minimum of the energy in $\C_0$ is the
vacuum, and in the sector $\C_1$ it is the 't Hooft--Polyakov monopole,
which is spherically symmetric \cite{tH1,Po}. Multi-monopole solutions in the 
higher sectors $\C_n$ are harder to find; however, they are understood rather 
well in the limiting version of the theory where the Higgs potential $V$ 
vanishes, but the asymptotic Higgs field still has a non-zero vacuum
expectation value. In this limit, multi-monopole solutions can be
found by solving first-order Bogomolny equations \cite{Bo,AH}. As for vortices
at critical coupling, there is a large moduli space of solutions (of
dimension $4n$) representing monopoles at arbitrary locations \cite{We1}. The
location of each monopole accounts for three of the four moduli per
monopole. The fourth is a phase that (for one monopole) appears to be
a gauge artifact, but it is now well understood that relative phases
between monopoles are physical, and even for one monopole, a
time-dependent phase is what can give the monopole an electric charge,
turning the monopole into a dyon. (Note, the moduli are not simply
positions and phases when the monopoles are close together and merge,
but the moduli space remains smooth.)

When the Higgs potential $V$ is positive, and the Higgs field massive, 
there are no Bogomolny equations, so it is necessary to solve the full 
second-order field equations. Fewer solutions are known in this case, 
but the basic monopole and the analogue of the most compact of the 
Bogomolny multi-monopole solutions persist. In particular, a 
2-monopole with toroidal symmetry persists \cite{KKT}. However, the basic 
monopoles tend to repel each other, making this 2-monopole unstable. It is
therefore a sphaleron.

The most interesting sphaleron involving monopoles was proved to exist 
by Taubes \cite{Tau}. Taubes considered monopole-antimonopole 
configurations in the vacuum sector $\C_0$ of the Georgi--Glashow 
model. Such a pair usually annihilate, but because a
monopole has a phase, it is possible to construct a closed loop of
monopole-antimonopole configurations where the relative phase
increases by $2\pi$. When the relative phase is $0$ or $2\pi$, or
indeed any value other than $\pi$, the configuration can simply collapse
to the vacuum, but when the phase is $\pi$ the monopole and
antimonopole are in an unstable balance. The minmax point along loops
of this type is a sphaleron solution, and is unstable to a 
change of phase away from $\pi$, in either direction. It can be
interpreted as a slightly distorted monopole-antimonopole pair at 
finite separation. The detailed (axisymmetric) solution has been 
found by careful numerical work \cite{Rub,KlKu,SV}. The energy is confirmed
to be less than that of a monopole and antimonopole at infinite 
separation, as anticipated by Taubes' general reasoning; the
magnetic dipole moment has also been calculated.  

Taubes' original argument for the existence of the
monopole-antimonopole sphaleron was mainly topological, although combined
with careful analysis. The sector $\C_0$ is topologically the space 
of maps $\Phi_{\infty}: S^2 \to S^2$ with degree 0. This space has 
non-contractible loops because its first homotopy group is
\be
\Pi_1({\rm Maps}(S^2 \to S^2)) = \Pi_3(S^2) = \Z \,,
\ee
where $\Pi_3(S^2)$ denotes the third homotopy group of $S^2$.
This result is a special case of a very useful, more general result
for based maps,
\be
\Pi_k({\rm Maps}(S^l \to M)) = \Pi_{k+l}(M) \,.
\ee

Taubes' work on the monopole-antimonopole solution was the inspiration
behind the discovery by this author, in collaboration with
F. Klinkhamer, of the electroweak sphaleron \cite{Ma9,KlMa}.
In the electroweak theory, the asymptotic Higgs field (in radial
gauge) defines a map $\phi_{\infty}: S^2 \to S^3$. The space of such maps is
connected, so the field configuration space $\C$ has just one 
component. From a topological perspective, therefore, 
there are no topological soliton charges in the electroweak theory, and in
particular, no smooth, finite-energy monopole solutions are expected.

However, $\C$ has non-contractible loops, because   
\be
\Pi_1({\rm Maps}(S^2 \to S^3)) = \Pi_3(S^3) = \Z \,.
\ee
By applying the minmax argument to such loops, and by more detailed
work constructing fields and solving the field equations, it was
possible to find an explicit sphaleron solution in the electroweak
theory. Higher-energy sphalerons with more unstable modes have also
been found, by considering non-contractible spheres of electroweak
fields \cite{Kli2}. 

\vspace{2mm}

\section{Properties of the Electroweak Sphaleron}
\vspace{3mm}

A limiting version of electroweak theory is where the weak mixing
angle vanishes. This is the limit where the U(1) field decouples and
the masses of the $W$ and $Z$ bosons are equal. The field
equations reduce to those of an SU(2) gauge field coupled to a complex
Higgs doublet, and the sphaleron is spherically symmetric in this
case. The version of electroweak theory that is realised in nature is
not far from this limit. The observed weak mixing angle is less than $30^\circ$
and the $Z$ boson mass is about 91 GeV, not much larger than 
80 GeV, the mass of the $W$ bosons. In the full U(2) electroweak 
theory the sphaleron is axisymmetric, and has a magnetic dipole moment 
\cite{KlMa,Jam}. Recall that the unbroken gauge group is the 
electromagnetic U(1), so the long-range field of a static solution 
is purely magnetic.

The energy of the sphaleron depends on the Higgs boson mass
and on the weak mixing angle. When the sphaleron solution was 
originally discovered, the Higgs boson had not been observed, and its
mass was poorly constrained. The sphaleron energy was then
estimated to be somewhere in the range 8 -- 14 TeV. Now that the Higgs boson
is known to have a mass of 125 GeV, a little more than the mass of
the $Z$ boson, the sphaleron energy is estimated to be approximately
9 TeV. This assumes that one can rely purely on the classical field
equations, combined with the experimentally determined coupling and
mass parameters. The contribution of the magnetic dipole field to the
energy is only about $1\%$.

A Higgs boson mass of 125 GeV is rather small from one
perspective. The sphaleron solution for this mass has $\phi = 0$ at
its centre, but it has been shown that for very large Higgs mass,
of order 1 TeV or higher, the sphaleron solution has a broken
discrete symmetry, and the field value $\phi = 0$ is not attained 
at any point in space \cite{BK,Ya}. The solution is closer to 
$|\phi| = 1$ everywhere, and in the limit of infinite Higgs boson mass, 
the electroweak sphaleron becomes a gauged Skyrmion \cite{EKS}. This 
exists in an effective field theory where the Higgs field is
constrained to its vacuum manifold $S^3$, changing the topology of the 
theory, but the relatively recent observation of the Higgs boson rules 
this effective field theory out. 
 
The sphaleron energy density is remarkably high. The length scale of 
the solution is the inverse of the masses of the contributing gauge and Higgs
fields, of order $(100 \, {\rm GeV})^{-1}$. This is approximately
$10^{-17}$ m, about 100 times smaller than the length scale of a proton.
The sphaleron volume is therefore about $10^6$ times smaller than that
of a proton. As the sphaleron energy is about $10^4$ times the mass of a
proton, its energy density is about $10^{10}$ times that of
a proton at rest. This, by itself, suggests the sphaleron is hard to
produce. 

Such energy densities appear to be unreachable in collisions at the
LHC -- CERN's Large Hadron Collider. There, colliding protons each have an
energy of more than 6 TeV (let's optimistically call this 10 TeV) 
and they are Lorentz contracted in the centre of mass frame by a factor
of $10^4$, the ratio of 10 TeV to the proton mass of 1 GeV. The
energy density is therefore $10^8$ times that of a proton at rest, and
it is in the form of a rather thin pancake, as there is no transverse
Lorentz contraction. This does not appear to be enough to produce 
sphalerons, although there
are millions of collisions per second, and large fluctuations of the energy
density must sometimes occur. Even if the energy density were two orders of
magnitude larger, it could be hard to produce a sphaleron as the
field energy, mainly in the form of quarks and gluons, would have to
transfer into a coherent combination of $W$, $Z$ and Higgs fields. 
Such a field can be interpreted as a coherent combination of about 10 
each of $W^\pm$, $Z$ and Higgs particles. Therefore, the non-perturbative 
process of sphaleron production in particle collisions is generally 
thought to be exponentially suppressed, in the same way that 
soliton-antisoliton production is suppressed \cite{DL}. However,
the production rate may be enhanced if a strong magnetic field is
present, in a region comparable to the sphaleron size. And the
production rate is almost certainly enhanced at high temperatures \cite{GRS}. 

Whether production of a sphaleron in particle collisions is at all likely may 
become clearer when experiments at LHC, or at somewhat higher energy,
find evidence for simultaneous production of two or more Higgs particles
together with a few $W$ or $Z$ bosons. The signal for this would be the
production of several high-energy leptons (electrons, muons or neutrinos). 

Remarkably, sphaleron production and decay is 
associated with a net change in baryon number $B$ and lepton number 
$L$ \cite{tH2}. This is the result of an anomaly in
baryon and lepton number conservation laws, and related to the fact
that a sphaleron has Chern--Simons number $\half$. Sphaleron production 
is therefore potentially extremely important, as it may help us 
understand the baryon asymmetry of the universe. The universe is 
dominated by matter (protons) rather than antimatter (antiprotons) 
but the source of the asymmetry remains unknown (although there are 
many ideas). Certainly, any observation of baryon or lepton number 
violation would be revolutionary, as no experiment so far has ever 
detected such a violation. However, measuring a net change in
baryon number in a high-energy collision may be hard, as many mesons 
and baryons, and also antibaryons, are produced. It may be easier to
keep track of charged leptons, but the neutrinos carry lepton
number too, and are generally undetected.  

\vspace{2mm}

\section{Sphalerons in the Skyrme model}
\vspace{3mm}

The Skyrme model is a field theory of mesons, an effective theory of the
strong interactions at relatively low energy, in which the quarks and gluons of 
QCD have been eliminated \cite{Sk,book,RZ}. It has topological
solitons -- Skyrmions -- that provide models for nucleons (protons and
neutrons) and all larger atomic nuclei. 

The basic Skyrme model just 
has interacting pion fields, the three pion particles being the lightest 
mesons, but variants have additional fields representing rho and/or 
omega mesons. In detail, the Skyrme model is a 3D sigma model, meaning 
that its field $U({\bf x})$, at a given time, is a map from $\R^3$ to 
a target space SU(2). An SU(2) matrix $U({\bf x})$ can be written as 
\be
U({\bf x}) = \sigma({\bf x}) 1_2 + i (\pi_1({\bf x})\tau_1 + 
\pi_2({\bf x})\tau_2 + \pi_3({\bf x})\tau_3) \,,
\ee
where $\tau_i$ are the three Pauli matrices, and the constraint 
$\sigma^2 + \pi_1^2 + \pi_2^2 + \pi_3^2 = 1$ needs to be imposed.
This links the Skyrme field $U$ with pion fields $\pi_1,
\pi_2, \pi_3$ and explains the name `sigma model'. 
The energy of a (static) Skyrme field is an integral of terms mainly 
involving the gradient of $U$, with an additional term
proportional to $1 - \sigma({\bf x})$ that accounts for the small mass of pions.

The group SU(2), regarded as a manifold, is the 3-sphere $S^3$, as is
implied by the above constraint. For the energy of a Skyrme field to
be finite, the boundary condition $U \to 1_2$ as
$|{\bf x}| \to \infty$ must be satisfied. This boundary condition 
allows a topological compactification of space to a 3-sphere, by
adding a point at infinity. A Skyrme field therefore becomes a based map 
$U: S^3 \to S^3$ (the base point condition being that the point at 
infinity maps to the unit element of SU(2)).

A map $U: S^3 \to S^3$ is characterised topologically by its degree, 
an integer $B$. The configuration space of the Skyrme model
is therefore disconnected, with its component $\C_B$ being the sector with
baryon number $B$. The notation $B$ rather than $n$ is used, as $B$ is 
identified with baryon number. This was Skyrme's great insight -- that 
a nonlinear theory of interacting pions could also account for
baryons, by treating baryon number as topological. 

In any continuous field evolution, the baryon
number does not change in the basic Skyrme model. However, when the
Skyrme model is extended to include electroweak fields then the
topological structure is more complicated, and the baryon number and
lepton number violation induced by electroweak sphalerons that we
discussed in the last section are probably possible. I'm not sure if
the details of this have been worked out, but a related phenomenon
in the context of Skyrmions is the expectation of baryon and lepton number 
violation when Skyrmions interact with monopoles \cite{Ch}. 

The various components $\C_B$ of the Skyrme model have been much 
studied. For many values of $B$, the minimal energy solution has been
found, and is known as the Skyrmion with baryon number $B$. 
Finding this requires considerable numerical work, but
various analytical and geometric ideas have been helpful to set up
initial configurations close to true 
solutions. It appears that in the basic Skyrme model, the global 
minimum of the energy is attained in each sector $\C_B$, but there is 
no proof of this. Calculations suggest that it is not favourable 
for a Skyrme field of baryon number $B$ to break up into 
infinitely-separated subclusters whose baryon numbers add up to $B$, 
but these calculations have loopholes \cite{MSS,Sch}.

\begin{figure}[h]
\centering\scalebox{1.0}{\includegraphics[width=0.7\textwidth]
{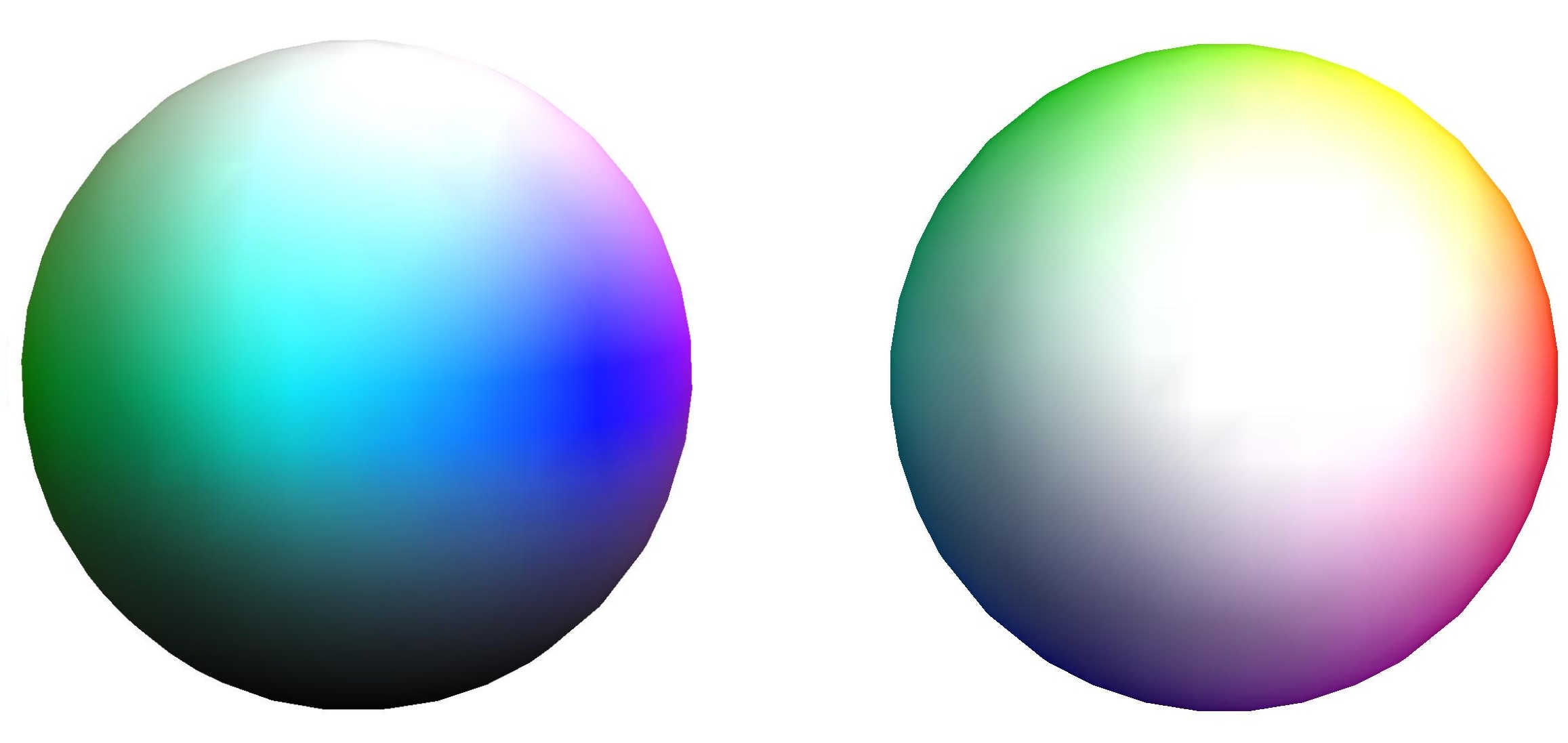}}
\vskip 0pt
\caption{$B=1$ Skyrmion (two different orientations)}
\end{figure}

The known Skyrmions with baryon numbers $B = 1,2,3$ and $4$ have,
respectively, spherical, toroidal, tetrahedral and cubic symmetries \cite{BTC}. 
The solutions with $B=1$ and $B=4$ are shown in Figs. 1 and 2. A
surface of constant energy density is displayed, and the colouring
indicates which pion fields have dominant values close to $1$ or $-1$. The 
coupling parameters of the Skyrme model are chosen so that 
the $B=1$ Skyrmion has an energy scale and length scale comparable to those 
of a nucleon. However the classical Skyrmion by 
itself does not model such a particle. For this it is necessary to 
quantize the rotational motion of the Skyrmion to obtain spin $\half$ 
states \cite{ANW}. The possibility of spin $\half$ arises from topology. The 
relevant space of maps has a non-trivial first homotopy group 
\be
\Pi_1({\rm Maps}(S^3 \to S^3)) = \Pi_4(S^3) = \Z_2 \,,
\ee
so there are non-contractible loops. In particular, it has been shown
that a rotation of a $B=1$ Skyrmion by $2\pi$ is such a loop. The quantum 
theory therefore allows the Skyrmion's wavefunction to acquire a sign 
change under the $2\pi$ rotation; the wavefunction is single-valued only 
on the universal, double cover of the the field configuration space 
\cite{FR}. That is why the quantized $B=1$ Skyrmion can represent 
a spin $\half$ proton or neutron. The proton and neutron are
distinguished by their isospin state, which arises from the quantized 
isospin symmetry of the Skyrme model that rotates the three pion 
fields among themselves.

\begin{figure}[h]
\centering\scalebox{0.7}{\includegraphics[width=0.7\textwidth]
{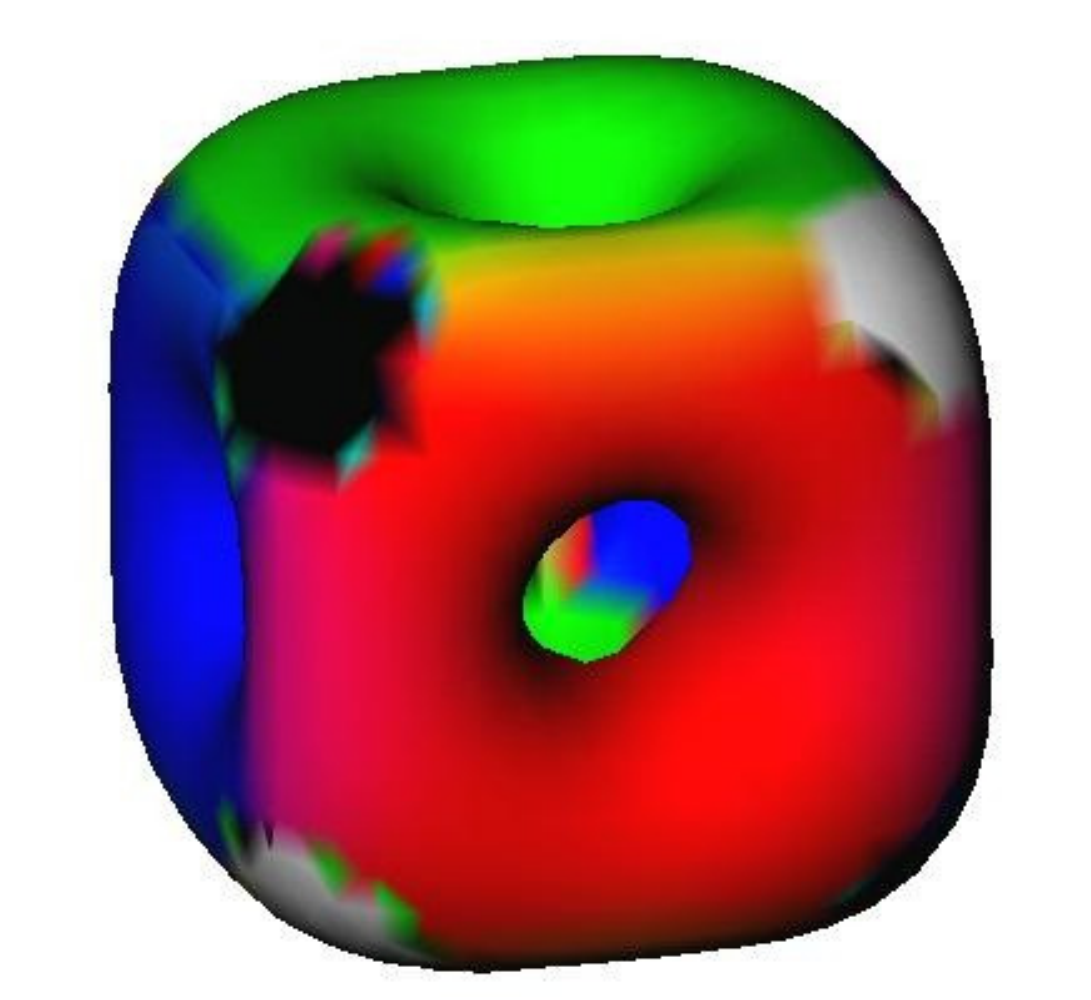}}
\vskip 0pt
\caption{$B=4$ Skyrmion}
\end{figure}

The topology of the space of maps $U: S^3 \to S^3$ is quite rich, so within
each sector $\C_B$ there are non-contractible loops and spheres of
various dimensions, and as a result, there are numerous unstable solutions of
the field equations, i.e. sphalerons in the Skyrme model. Some of
these have quite low energy, only slightly higher than the minimal
energy Skyrmion in that sector. It is not possible to describe all 
solutions of this type systematically, so we just mention a few that are known.

For each positive baryon number, and not just $B=1$, there is a
spherically symmetric solution \cite{JRh}. The field has an angular dependence
like the $B=1$ solution, accompanied by a radial profile function that winds
$B$ times between the field values $U=1_2$ and $U=-1_2$. The structure is
onion-like, and this is not at all energetically favourable, so apart
from the $B=1$ solution, all these solutions are unstable. For
example, in the limit of massless pions, the spherically symmetric 
$B=2$ solution has energy 2.98 times that of the $B=1$ solution, and 
has six unstable modes, whereas the 
toroidal $B=2$ solution has energy 1.91 times that of the $B=1$
solution, and is stable. 

There was an attempt to use the unstable manifold of the spherically
symmetric solution (defined by following the curves of
steepest descent of the energy) as a landscape of $B=2$ field
configurations, but there are technical and numerical difficulties with 
this \cite{Ma4,AM2}. To connect the Skyrme model with nuclear physics, it is
desirable to find finite-dimensional configuration spaces of Skyrme fields to
quantize, for each $B$. Rigid-body quantization of Skyrmions (the 
stable minima of the Skyrme energy) is the best known approach to 
quantization \cite{ANW}, but imposes too strong constraints on the dynamics. 
States of nuclei do not simply form a single band of rotational excitations 
of one underlying structure. At the very least, vibrations away from the 
minimal energy Skyrmion should be included \cite{GH}, and quantized, but to do 
this in a consistent nonlinear way is not easy. Using the unstable manifolds 
of saddle point solutions still appears to be an attractive route towards a 
systematic approach to quantization.

There is a large class of unstable, sphaleron solutions in the vacuum sector of
the Skyrme model, with $B=0$. Some of these are Skyrmion-antiSkyrmion
configurations in unstable equilibrium \cite{KrSu,ShTc}, analogous to
the monopole-antimonopole solution of Taubes. An alternative construction
is to select an equatorial 2-sphere of
the target $S^3$, and seek solutions whose values lie everywhere in
this 2-sphere. The Skyrme model with target restricted to $S^2$ is the
Skyrme--Faddeev model, known to have many solutions with a
knot-like character, called Hopfions. These solutions may have some 
interpretation within the purely mesonic or gluonic sector of strong 
interactions, for example as glueballs, but further investigation of 
their role is needed. All these solutions are unstable within the
Skyrme model, as the target $S^2$ can be slipped off the equator of 
$S^3$, making it smaller and reducing the field energy \cite{Fo}.

\begin{figure}[h]
\centering\scalebox{0.8}{\includegraphics[width=0.7\textwidth]
{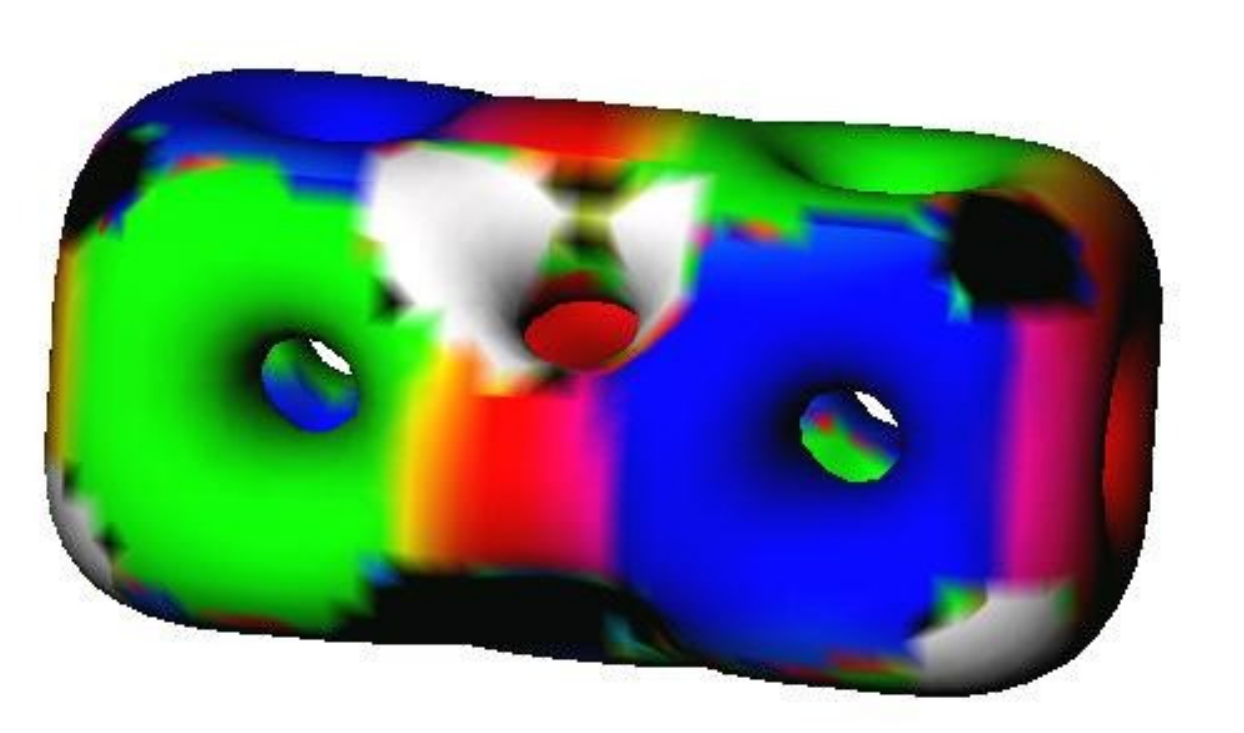}}
\vskip 0pt
\caption{$B=8$ Skyrmion}
\end{figure}

For higher $B$, above about $B=7$, it is found that the landscape of
local minima in the Skyrme model becomes quite complicated, and it
is hard to determine numerically which of the local minima is the global
minimum. The energies of these minima are very close, and it doesn't 
really matter which is the global minimum (the true Skyrmion), as a 
successful quantization should incorporate all the fields with energy 
close to the minimum. Between these local minima there are inevitably 
saddle point solutions, and some of these have energies only a little 
higher than the local minima.

\begin{figure}[h]
\centering\scalebox{0.6}{\includegraphics[width=0.7\textwidth]
{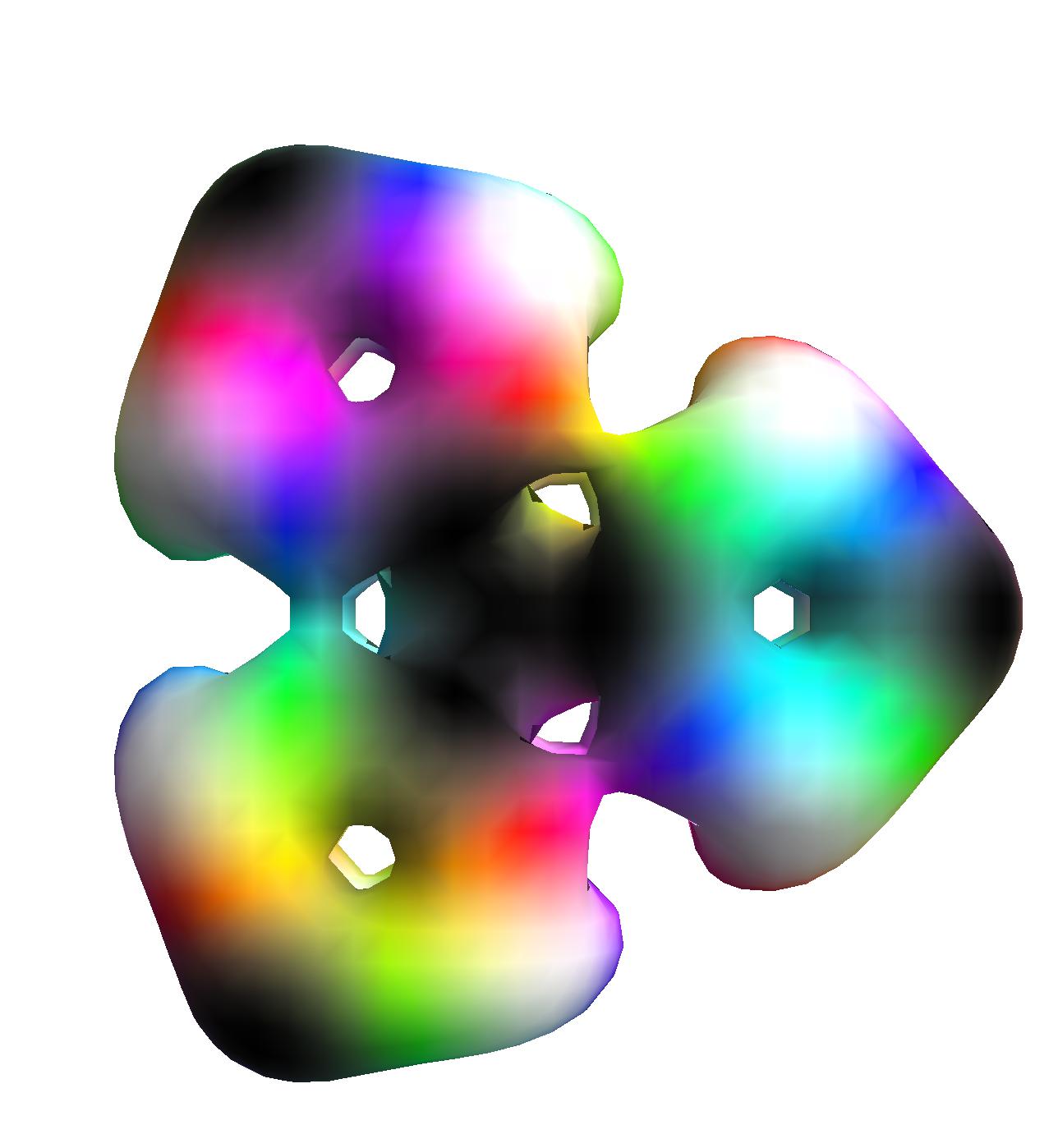}}
\vskip 10pt
\caption{$B=12$ Skyrmion with $D_{3h}$ symmetry}
\end{figure}

An example occurs for $B=8$ \cite{BMS,Fei}. One solution is shown in 
Fig. 3 and this is the global minimum of the energy when the pion 
mass is close to its physical value. Its rigid-body quantization gives
a reasonable model for the Beryllium-8 nucleus, but is less successful for the
isobars Lithium-8 and Boron-8 \cite{BMSW}. The solution is clearly made
of two $B=4$ cubic Skyrmions brought close together. In the $B=8$ 
solution, the merged faces of the cubes have the same colour (red in 
the figure), because this is what leads to an attraction. However, 
it is still possible to twist one cube relative to the other around
the line joining them, and this takes little energy. So there is a solution
similar to that in Fig. 3 where the two cubes have the same
orientation (both nearby faces are green or blue, rather than one green and
one blue), but this is slightly unstable to untwisting \cite{GH}. 
Rotating one cube relative to the other by $\pi$ creates a
closed loop of field configurations along which there is one minimum of 
the energy and one saddle point. A good quantization should take into 
account the degree of freedom along the loop, in addition to all 
the rigid-body degrees of freedom (translations, rotations, isospin 
rotations). This would give a better model of Beryllium-8 and its
isobars than simply rigid-body quantization.     

\begin{figure}[h]
\centering\scalebox{0.8}{\includegraphics[width=0.7\textwidth]
{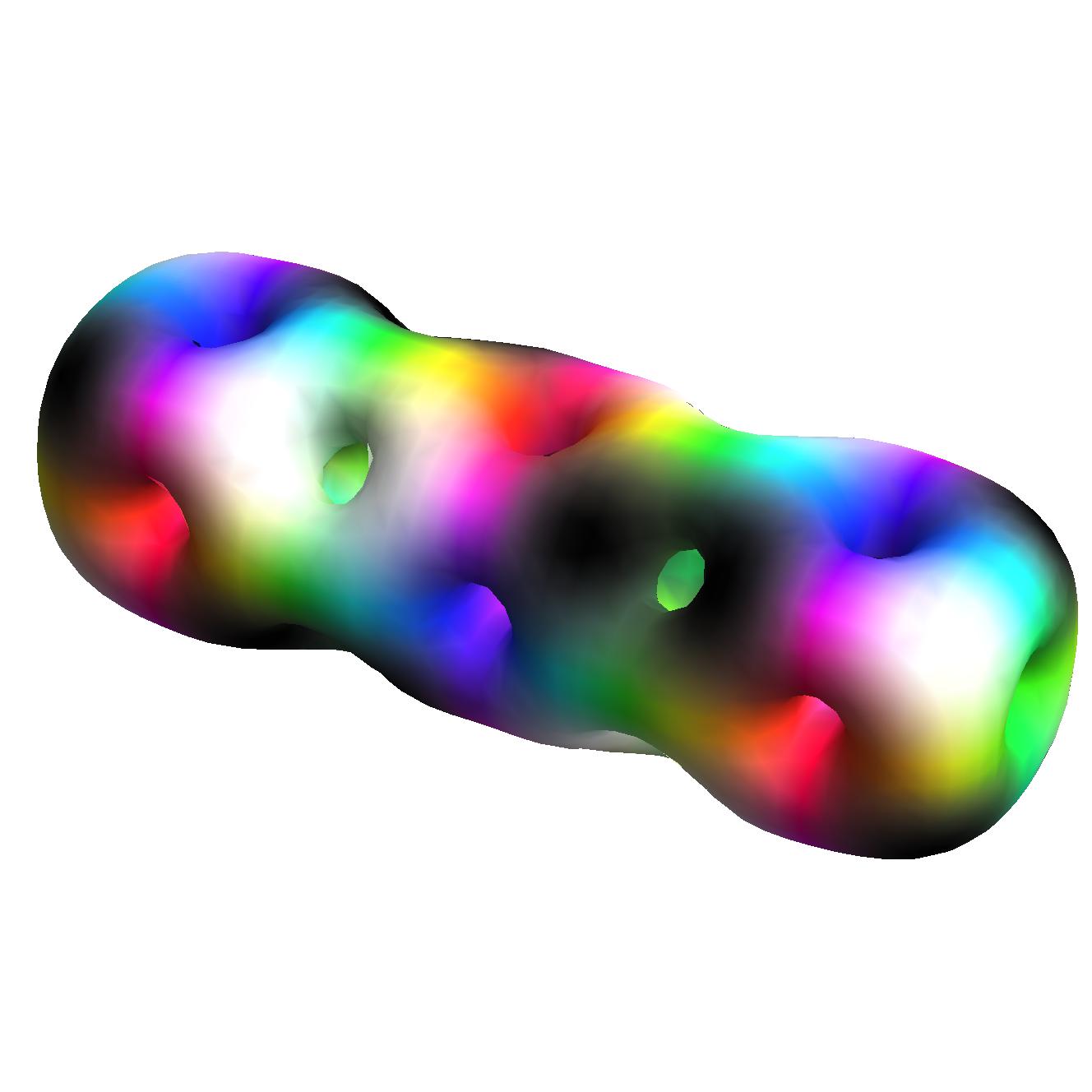}}
\vskip 0pt
\caption{$B=12$ Skyrmion with $D_{4h}$ symmetry}
\end{figure}

As a final example, we show two solutions with $B=12$ in Figs. 4 and
5 \cite{BMS,Lau}. These consist of three $B=4$ cubes arranged to
attract, forming an equilateral triangle or a straight chain. Again 
their energies are very similar, and it is not clear which is the true
Skyrmion. It is possible that one shape is stable and the other is a 
saddle point, but it is also possible for both to be local minima,
with a saddle point (of L-shape) in between. These different scenarios 
are hard to distinguish, and are sensitive to the exact value of the 
pion mass, and which variant of the Skyrme model one works with. 
Quantization of these configurations separately (treating 
them as well separated by an energy barrier) has given a moderately 
good understanding of the ground and excited states of 
Carbon-12 \cite{LM}, including the Hoyle state, but a better model 
is obtained by taking account of the landscape between these 
configurations \cite{Raw}.  

\vspace{2mm}

\section{Conclusions}
\vspace{3mm}

Unstable, saddle points of the energy occur frequently in
field theory -- they are called sphaleron solutions, in
contrast to the stable solitons that may also exist.
Sphalerons often have a topological interpretation, and are
related to non-contractible loops or spheres in field configuration
space $\C$. A saddle point is also expected to occur
between any pair of local energy minima in a connected
component of $\C$.

Gauge-Higgs theories in 3D are good sources of sphalerons, as well as 
solitons, as these theories have nontrivial topology captured by
the Higgs field at infinity (in radial gauge) and the solutions 
have a preferred, finite length scale.

The electroweak sphaleron is one such solution. Its energy of 9 TeV
has now been reached in proton collisions at the LHC, but the sphaleron's 
very high energy density, and its coherent electroweak field structure make 
it hard to produce.

There has been an interesting suggestion (discussed at this Royal
Society meeting) that it could be easier to create a sphaleron in a 
strong magnetic field. Some of the strongest known fields arise
briefly in heavy ion collisions at the LHC. The effective energy of 
an electroweak sphaleron will be lower in a strong magnetic field, 
whenever its magnetic dipole moment is aligned with the field.

\vspace{2mm}

\section*{Acknowledgements}

I am grateful to the organisers of this Royal Society Scientific Discussion
Meeting: Topological Avatars of New Physics, and especially Prof. N. 
Mavromatos, for the invitation to speak. Some of the points in the
concluding section arose from the discussion session, and I acknowledge the
contributions of several participants.

\end{document}